\pdfoutput=1
\documentclass[journal]{IEEEtran}
\usepackage{cite}
\usepackage{amsmath,amssymb,amsfonts}
\usepackage{algorithmic}
\usepackage{graphicx}
\usepackage{textcomp}
\usepackage{graphicx}
\usepackage{bm}

\usepackage{mathrsfs}
\usepackage{stfloats}

\usepackage{array}
\usepackage{url}
\usepackage[colorlinks,linkcolor=black,anchorcolor=black,citecolor=black]{hyperref}
\usepackage{booktabs}
\usepackage{tabularx}
\usepackage{multirow}
\usepackage{diagbox}
\usepackage{booktabs}
\usepackage{color}
\usepackage{bbding}
\def\BibTeX{{\rm B\kern-.05em{\sc i\kern-.025em b}\kern-.08em
    T\kern-.1667em\lower.7ex\hbox{E}\kern-.125emX}}

\begin{document}

\title{3D Brain Reconstruction by Hierarchical Shape-Perception Network from a Single Incomplete Image}
\author{Bowen Hu,
Baiying Lei,
Shuqiang Wang,
Yong Liu,
Bingchuan Wang,
Min Gan,
Yanyan Shen
\thanks{This work was supported by the National Natural Science Foundations of China under Grant 61872351}
\thanks{Bowen Hu, Shuqiang Wang and Yanyan Shen are with the Shenzhen Institutes of Advanced Technology, Chinese Academy of Sciences, Shenzhen 518055, China. Bowen Hu is also with University of Chinese Academy of Sciences, Beijing 100864, China, Email: sq.wang@siat.ac.cn}
\thanks{Baiying Lei is with the School of Biomedical Engineering, Shenzhen University, Shenzhen 518060, China}
\thanks{Yong Liu is with the Gaoling School of Artificial Intelligence, Renmin University of China, Beijing 100872, China}
\thanks{Bingchuan Wang is with the School of Automation,Central South University, Changsha 410083, China}
\thanks{Min Gan is with the College of Computer Science and Technology, Qingdao University, Qingdao 266071, China}
}
\maketitle

\begin{abstract}
3D shape reconstruction is essential in the navigation of minimally-invasive and auto robot-guided surgeries whose operating environments are indirect and narrow, and there have been some works that focused on reconstructing the 3D shape of the surgical organ through limited 2D information available. However, the lack and incompleteness of such information caused by intraoperative emergencies (such as bleeding) and risk control conditions have not been considered. In this paper, a novel hierarchical shape-perception network (HSPN) is proposed to reconstruct the 3D point clouds (PCs) of specific brains from one single incomplete image with low latency. A branching predictor and several hierarchical attention pipelines are constructed to generate point clouds that accurately describe the incomplete images and then complete these point clouds with high quality. Meanwhile, attention gate blocks (AGBs) are designed to efficiently aggregate geometric local features of incomplete PCs transmitted by hierarchical attention pipelines and internal features of reconstructing point clouds. With the proposed HSPN, 3D shape perception and completion can be achieved spontaneously. Comprehensive results measured by Chamfer distance and PC-to-PC error demonstrate that the performance of the proposed HSPN outperforms other competitive methods in terms of qualitative displays, quantitative experiment, and classification evaluation.
\end{abstract}

\begin{IEEEkeywords}
Shape reconstruction, hierarchical shape-perception, attention gate block, point cloud.
\end{IEEEkeywords}

\IEEEpeerreviewmaketitle

\section{Introduction}

\IEEEPARstart{m}{inimally-invasive} and auto robot-guided surgeries have gradually been used in brain surgery, bringing patients more minor surgical wounds, shorter recovery time, and better treatment experience. Because of the new visual environment and navigation methods of these surgeries, new requirements for the acquisition ability of intraoperative information are put forward. Since doctors cannot directly observe lesions and surgical targets during operations, their experience is often not so efficient. Recently, the application of intraoperative MRI (iMRI) has become more and more extensive, and some works used it to relieve the stricter visual restrictions in minimally invasive surgery \cite{van2018open,thomas2017novel}. But unlike the rich internal details of the brain, 2D MRI cannot provide intuitive and visually acceptable information of the surface and the shape of target brains that is more important for surgery. Because the space complexity of computing 3D MRIs directly is $O(n^3)$, it is also not advisable to use it to perceive the shape of the target brain in operations that require real-time algorithms. What's more, since the 3D MRI representation can not correlate with the coordinate position of brains directly, doctors must manually assist in surgery navigation, thus reducing the automaticity of the surgery navigation system. These two facts jointly result in the scarcity of the visual support of brain minimally-invasive and auto robot-guided surgeries. Thus, it is a necessary direction of development for these types of surgery to find some indirect 3D shape information acquisition methods that are accurate and controllable. Furthermore, because of the limitations of conventional scanners and medical environments, these methods should reduce reliance on physical equipment and requirements for traditional information.


There have been some works that focus on 3D shape reconstruction \cite{35,36} from images to help doctors obtain additional visual information. PCs are used as the representation of the reconstruction result. A point cloud, which is represented as a set of points in 3D space, uses N vertices to describe the target. In addition to providing an accurate target shape, PC representation carries local position coordinate information at each point, which can be used for automated navigation of medical robots and devices. Therefore, it is a reasonable proposal to continue choosing PCs as the reconstruction representation in brain surgical scenes. However, most existing methods suffer from at least one of the following two problems. The first being that They ignore that images obtained for shape reconstruction are often incomplete and damaged Limited by the light environments of optical sensing devices and various possible visual pollution (such as local bleeding) outside of the surgical plan. And the second being that their requirements are too strict about the number and angle of input images. Reconstruction methods with too many inputs may lead to longer processing times, thus increasing the risk of intraoperative accidents. Up to now, no effort has been devoted to solving both of these existing problems at once in 3D point cloud reconstruction while maintaining accuracy. The main purpose of this paper is to find a method that can perceive and reconstruct the shape of targets from as few potentially incomplete images as possible.

In order to reconstruct the accurate and complete PC structure from the incomplete image from a single incomplete image, a novel and composite model based on generative adversarial architecture and multilayered encoder-decoder structure named hierarchical shape-perception network (HSPN) is proposed in this paper to efficiently complete 3D shape reconstruction tasks and meet the specific needs of brain surgery scenarios as much as possible. The encoder of HSPN consists of a predictor based on Generative adversarial network (GAN) architecture and several PointNet++ \cite{14} encoding blocks, while the decoder consists of multilayered decoding blocks. Hierarchical attention pipelines that can transmit the extracted feature information are constructed between corresponding encoding and decoding blocks. A novel branching generator that contains multiple graph convolutional networks is built in predictor to generate incomplete point clouds from a single incomplete image accurately. The encoding blocks and their corresponding structural-consistency decoding blocks perceive the shape of the target hierarchically, reconstruct the complete point cloud, and ensure different reconstruction levels can be guided by the corresponding shape structure. Besides, a novel module named attention gate block (AGB) is designed to unify the attention computing of local encoding features transmitted by hierarchical attention pipelines and reconstructing self-attention computing.

The main contributions of this paper can be summarized as follows:



1) The hierarchical shape-perception network is proposed to fulfill the task of 3D shape reconstruction in restricted surgical visual environments. This is the first work to reconstructing a complete brain point cloud from a single incomplete MRI image. The proposed model designs a variety of mechanisms to ensure the accurate perception of the 3D shape.

2) An adversarial predictor is employed to reconstruct the incomplete point cloud that can describe the image details as much as possible. A novel generator via branching graph convolutional network (GCN) is constructed to describe the complex brain microstructure.

3) A level-to-level encoding and decoding system is designed to accurately perceive the shape of target brain subjects and restore incomplete point clouds. Hierarchical attention pipelines are constructed to transfer the local geometric features aggregated by each layer of the encoder to the corresponding decoding block. AGBs are designed to unify the attention computing process in HSPN.


\section{Related work}
Recently, machine learning and deep learning technology has been popularized in medical image processing, and has been applied in maturity recognition \cite{wang2018automatic,wang2018skeletal,wu20183d,wang2019ensemble,lei2020skin,wang2020ensemble}, disease analysis \cite{wang2018bone,9130073,zeng2017ga}, cross-modal data supplement \cite{hu2020brain,hu2020medical,hu2021bidirectional}, image segmentation \cite{9122459} and other fields. Many deep learning reconstruction models, such as GANs \cite{li2021hausdorff,zhang2018sch,8764602,hu2021point,yu2021tensorizing,wang2020diabetic} and variational methods\cite{wang2007variational,wang2008variational,wang2009variational,mo2009variational}, are widely used in reconstructing 2D images. There are also many works that combine these methods with 3D data \cite{hackel2016contour,wu2016learning,zhou2018voxelnet,9238491}. 3D shape analysis is an active research field in recent years \cite{12,13,14,15,16,17}, where the studies lead to many branches. For examples, 3D classification and recognition methods \cite{18,19} use the characteristics of 3D shapes to remove the perceptual limitations in traditional classification tasks and improve the efficiency of the algorithm; 3D reconstruction methods \cite{wang2018pixel2mesh,zhou2018voxelnet,20,21} restore the 3D structure of the target from each source domain, providing researchers with multi-view information; From the perspective of geometry based \cite{22,23,24} and alignment based \cite{25,26,27}, the 3D completion method explores the algorithm framework for completing the incomplete 3D structure. Our method is composed of multiple frameworks to complete a complex 3D complemented shape reconstruction task. Works related to these frameworks can be further categorized according to the input and output form.

\subsection{Point Cloud Generation}
The representational learning of 3D shapes is a specific research direction. This direction tends to build on well-constructed point cloud datasets and conducts indicative or extensional research by reconstructing the data input as closely as possible. In this field, \cite{28} proposed an auto-encoder model to fold and recover the point clouds, then demonstrated the validity of the method by the classification experiments with other model reduction results. \cite{29} proposed a variational auto-encoder model to generate close point cloud results and then apply them to fracture detection and classification. \cite{30,31,32} proposed different GAN architectures to learn the mapping from Gaussian distribution to multiple classes of point cloud representations, so as to reconstruct point clouds in an unsupervised manner.

\subsection{Image-to-PC Reconstruction}
The demand for image-to-3D shape conversion is continuously growing with the development of the medical imaging field. Benefiting from the iterations of medical tools, the advantages of 3D data over traditional images in terms of positional information, perceptual information, and visual acceptability are increasingly noticed. From limited image data and point cloud ground truth, \cite{33} proposed a new loss function named geometric adversarial loss to reconstruct point clouds representation that better fits the overall shape of images. In \cite{34}, a deep neural network model composed of an encoder and a predictor is proposed. This model, which is named PointOutNet, predicts a 3D point cloud shape from a single RGB image. On the basis of this work, \cite{35} applied this model to the one-stage shape instantiation to reconstruct the right ventricle point cloud from a single 2D MRI image, thereby simplifying the two-stage method proposed by \cite{36}.

\begin{figure*}[ht]
\centering
\includegraphics[width=18cm]{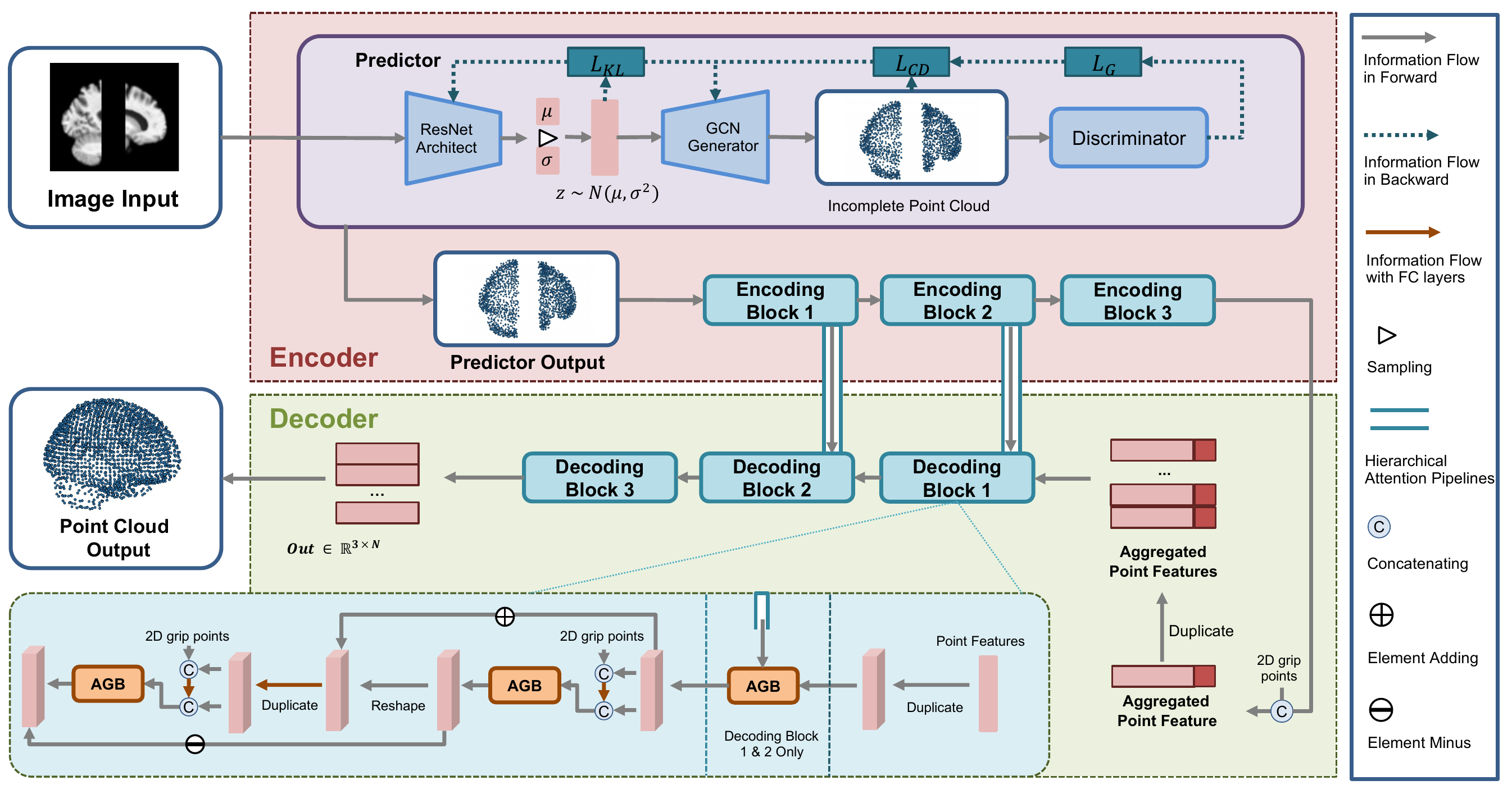}
\caption{The architecture of the proposed model. The pink area is the encoder network, composed of a predictor and three encoding blocks based on PointNet++. In addition to outputting the extracted features to the next block, the first two encoding blocks also feed them back to the corresponding decoding block through the attention pipeline. The green area represents the decoder network. The detailed structure of the decoding block is also given in the blue room.}
\end{figure*}

\subsection{Point Cloud Completion}
The 3D point cloud shape completion is a relatively complex area. However, due to the fact that a large amount of 3D data is facing incomplete and damage, researches in this area are also surging development. Based on PointNet \cite{18} and PointNet++ \cite{14}; PCN \cite{37}, FoldingNet \cite{28}, AtlasNet \cite{38} designed different encoder-decoder structures to extract global features from the target point cloud and restore the complete point cloud. In \cite{39}, a tree-structured network named TopNet was proposed to generate arbitrarily structured point clouds without explicitly enforcing a specific structure.

\section{Method}
The overall architecture of the proposed HSPN is shown in Fig. 1. HSPN consists of an encoder, which has a predictor and multiple encoding blocks, and a corresponding decoder. Considering the differences in the input and output of modules, we designed a unified information flow to communicate with adjacent modules and built independent hierarchical attention pipelines that transmit local attention features between corresponding encoding blocks and decoding blocks.

\subsection{AE Architecture with Hierarchical Encoder And Decoder}
\subsubsection{Encoder}
Generally, The universal AE framework aims to learn a generative model where the encoder aggregates a corresponding feature representation to describe the low dimensional generating factor. The representation can therefore be treated as the information flow that captures the necessary details about inputs. However, architectures of the standard AE are not ideal for this task as the distributions of image inputs and point cloud outputs are too differentiated. The completion task requires the encoder to provide more supplementary information for the decoder. Direct mapping conversion will result in confusing styles in different domains and losing generating details, which is intolerable for the medical point cloud generation. As basic requirements, the encoder needs to have the ability to convert a given image input into a point cloud as restored as possible and then describe its features in an appropriate representation.

To meet such demands, our encoder is designed into a special architecture consisting of a predictor and multilayered feature extraction blocks. As shown in Fig.1, the adversarial predictor is roughly a GAN framework, outputs point clouds $Y_{N \times 3}$ that can accurately represent the target brains (in this work, $N = 2048$). Then, multiple blocks composed by PointNet++ aims to extract the features from the incomplete point cloud. After several processes similar to the downsampling in CNN, the last block aggregates the point cloud into a latent feature, which carries the structure information of the input image. In particular, in order for the decoder to obtain enough information to restore the complete part of the point cloud and generate the incomplete part of the point cloud, each block will additionally share the sampling feature to the corresponding decoding block by hierarchical attention pipelines.

\subsubsection{Decoder}
In order to make the best use of the sampling features shared by the encoder, the composition networks of the decoder are considered to be designed as a one-to-one correspondence with multilayer encoding blocks. In this work, we build up hierarchical attention pipelines to share sampling features and set decoding blocks in the decoder to restore and generate the local microstructure of the target while completing a point cloud. We will introduce details in the following chapters.

\subsection{Adversarial branching GCN Predictor}
\begin{figure*}[htb]
\includegraphics[width=18cm]{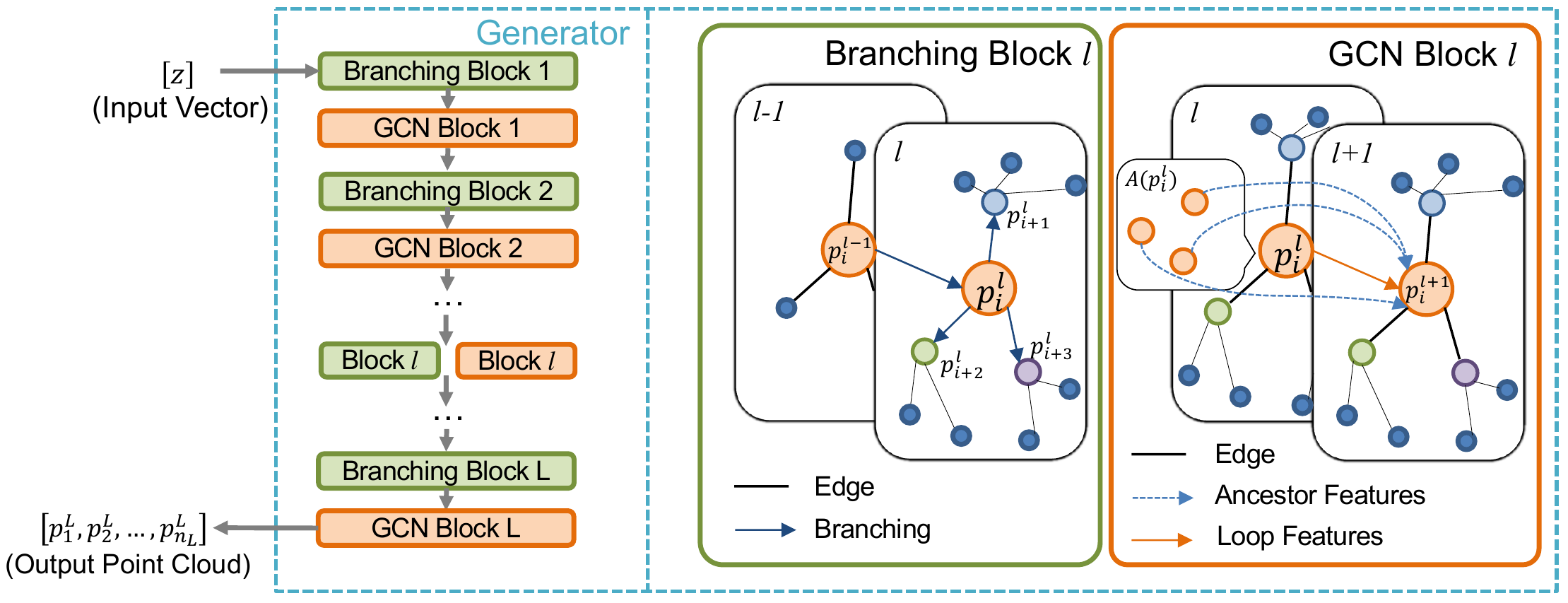}
\caption{The detailed structure of the generator in the predictor. An improved branching graph convolution network consisting of multiple GCN block and branching block groups is constructed.}
\end{figure*}
MRI images contain rich structural information. The first step of the completion is to restore the physical structure represented by these incomplete images. The adversarial predictor is designed to extract geometric information and use the information to reconstruct the point clouds as close to the original shape as possible. Given a specific brain subject, the point cloud of it can be expressed as a matrix $Y_{N\times3}$ which denotes a set of N points and each row vector represents the 3D coordinate of a vertex. Incomplete 2D images are required for the proposed adversarial predictor. The predictor has a ResNet \cite{he2016deep} model to take such images $I_{H\times W}$ as inputs, where H is the height and W is the width of an image. And the ResNet produces vectors $z \in \mathbb{R}^{96}$ from a Gaussian distribution with a specific mean $\mu$ and standard deviation $\sigma$ as outputs, which are treated as a point set with only one single point by the generator of the predictor. The generator has a series of GCN blocks and branching blocks to expand and adjust the initial point set. Then, a discriminator similar to WGAN-GP \cite{gulrajani2017improved} differentiates the output and real point cloud and then enhances the generator.

In order to accurately align the position of points to describe the complex structure of the brain, the implemented generator needs to have a strong perception of spatial shape. Because the dimensionality reduction operation of the fully connected layer destroys the potential contact information between adjacent points, and the efficiency of CNN is limited to traditional Euclidean data, GCN is considered to construct the generator. To adapt to the particularity of point cloud generation, multiple GCN block and branching block groups divide the brain into different brain areas and generate these areas respectively to enhance the generating details. As shown in Fig. 2, the GCN block is defined as
\begin{equation}
p^{l+1}_i = \sigma \left (\bm{F}^{l}_{K}(p^{l}_i) + \sum\limits _{q_j \in A(p^{l}_i)}U^l_jq_j + b^l\right ),
\end{equation}
where there are three main components: loop term $S^{l+1}_i$, ancestor term $A^{l+1}_i$ and bias $b^l$. $ \sigma (\cdot)$ is the activation function.

Loop term, whose expression is
\begin{equation}
S^{l+1}_i = \bm{F}^{l}_{K}(p^{l}_i),
\end{equation}
is designed to transfer the features of points to the next layer. Instead of using a single parameter matrix $W$ in conventional graph convolutional networks, the loop term uses a K-support fully connected layer $\bm{F}^{l}_{K}$ to represent a more accurate distribution. $\bm{F}^{l}_{K}$ has K nodes $(p^l_{i, 1}, p^l_{i, 2}, ..., p^l_{i, k})$, and can ensure the fitting similarity in the big graph.

The ancestor term allows features to be propagated from the ancestors of a vertex to the corresponding next connected vertex. This term of a graph node in conventional GCN is usually named neighbors term and uses the information of its neighbors, rather than its ancestors. But in this work, point clouds are generated dynamically from a single vector, so the connectivity of the computational graph is unknown. Therefore, this item is modified to
\begin{equation}
A^{l+1}_i = \sum\limits _{q_j \in A(p^{l}_i)}U^l_jq_j,
\end{equation}
to ensure structural information is inherited and multiple types of point clouds can be generated. $A(p^{l}_i)$ is the set of ancestors of a specific point $p^{l}_i$. these ancestors map features spaces from different layers to $p^{l}_i$ by using linear mapping matrix $U^l_j$ and aggregate information to $p^{l}_i$.

Branching is an upsampling process, mapping a single point to more points. Different branching degrees $(d_1, d_2, ..., d_n)$ are used in different branching blocks. Given a point $p^{l}_i$, the result of the branching is $d_l$ points. Therefore, after branching, the size of the point set becomes $d_l$ times the upper layer. By controlling the degrees, we ensure that the reconstruction output in this work is accurate 2048 points.

To ensure the accuracy and effectiveness of the reconstruction, it is reasonable to continuously transfer point features from the root to the bottom. Branching blocks maintain the information flow that achieves this process. The purpose of branching blocks is to build the tree structure of our generative model, thereby giving the generator ability to keep the relative position relationship between points and aggregate the prior shape knowledge in the ground truth in an unsupervised way.

\subsection{Encoding And Decoding Blocks}
After the predictor, we set multiple PointNet++ networks as encoding blocks to aggregate the shape information of point clouds into features. Except for the last encoding block, other blocks will present the aggregation results as intermediate features and transfers them to the next encoding block and the corresponding decoding block simultaneously. As a module to generate a global representation of the brain shape, the last encoding block merges all the features into a latent feature.

Then a hierarchical decoder is constructed to restore the original shape from the feature. Each layer of the decoder is a decoding block. The details are shown in Fig. 1. The decoding block is a complex network that uses fully connected layer networks as the skeleton and integrates various matrix operations and attention mechanisms. For the decoding blocks of the first two levels, we add an attention gate block to process the attention combination of global structural features and sampling features to ensure that the generation of each level focuses on the key decoding region.

\subsection{AGB: Attention Gate Blocks}
\begin{figure}[hb]
\includegraphics[width=8.8cm]{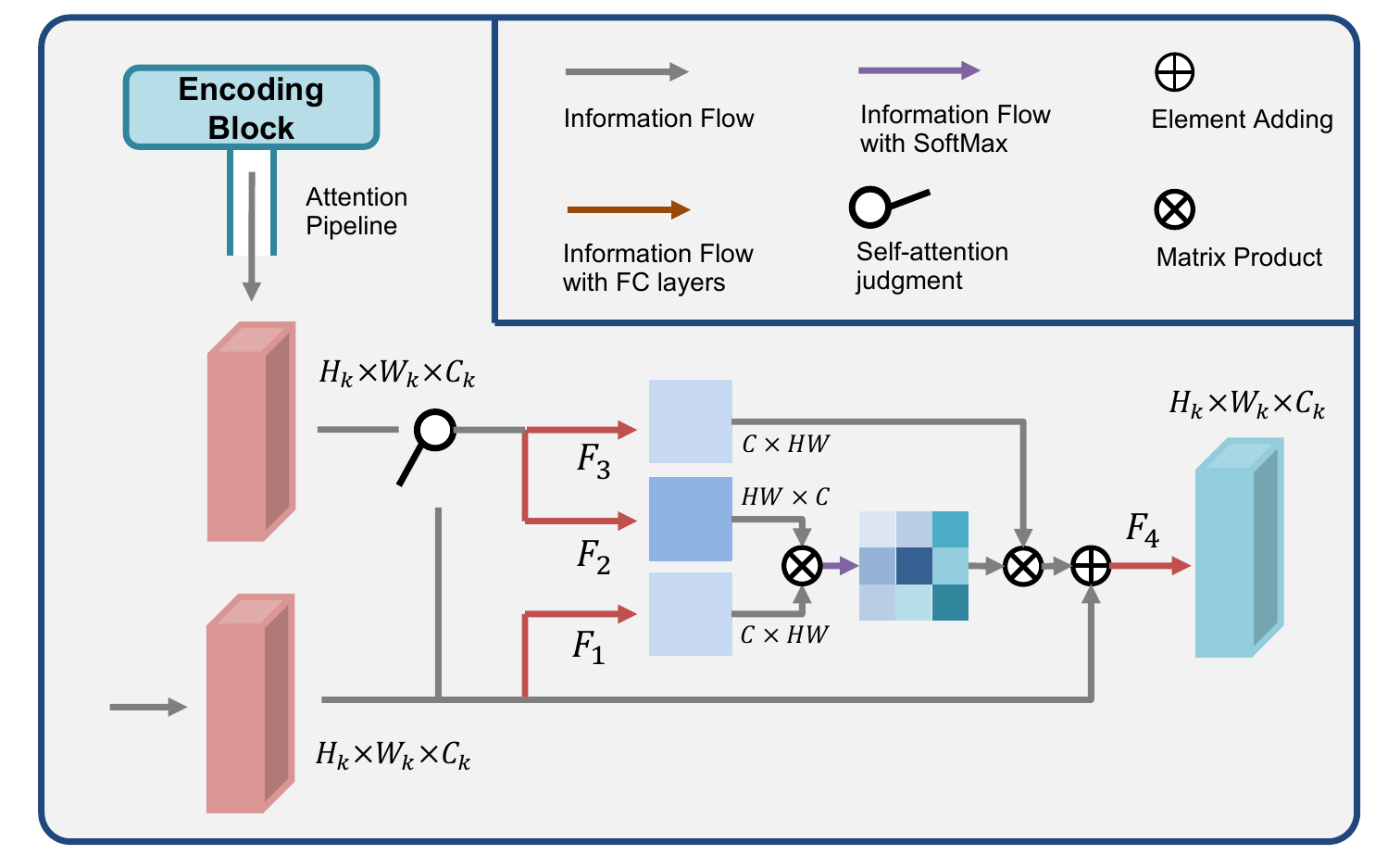}
\caption{The detailed structure of the proposed attention gate block. The self-attention judgment will judge whether there is an attention pipeline that transmits the local aggregated features of the point cloud to participate in the calculation. If not, the block will be converted into a self-attention module.}
\end{figure}

In the proposed HSPN architecture, there are two places where the attention mechanism is applied to enhance the fitting ability of the network. The first application is between the encoding blocks and their corresponding decoding blocks. We build hierarchical attention pipelines to pass the aggregated sampling features to the decoder and fuses them with the output of the previous layer. Through hierarchical attention pipelines, we hope that the entire model can better achieve two purposes. For the complete part of the brain point cloud that has been displayed by image information and restored by the predictor, pipelines should retain and transfer its microstructure as much as possible, and direct the decoder to generate such sub-regions more consistently. For the missing part of the point cloud corresponding to the missing area of the image due to various accidents or technical limitations, pipelines and attention processing networks should be able to extract features similar to the complete part of the PC based on symmetry or correlation information. These features will guide the decoder to complete the point cloud regions more physiologically reasonable to satisfy the pathological features of the same surgical case.

The second application is in the decoding module. We require that the decoding blocks have the ability to notice a preference of more different features in the input to generate point clouds, so it can maximize the use of the entire generation space and increase the generation efficiency. Therefore, we construct self-attention networks to integrate geometric information flow and distinguish more similar features.

In order to unify the application of these two attention networks and build a more generalized overall structure, we set up the attention gate blocks, whose details are shown in Figure 3. AGB accepts two inputs, and uses a self-attention judgment to judge whether there is an attention pipeline that transmits the local aggregated features of the encoding point cloud to participate in the calculation. When one of the two inputs is from the attention pipeline, AGB will be used as an attention network, using two fully connected layer networks to unify the two inputs and calculate the attention map. Otherwise, AGB will work as a self-attention network. Given two point sets, $P^l$ and $Q^l$ at the l-level of the decoder, as inputs of AGB, the AGB first calculates the attention scores between each point $p^l_i \in P^l$. For any $q^l_j\in Q^l$, the attention score $a^l_{ij}$ is computed as
\begin{equation}
a^l_{ij} = \frac{exp(\bm{F}^{l}_1(p^l_i)^T\cdot\bm{F}^{l}_2(q^l_j))}{\sum\limits _{q^l_k\in Q^l}exp(\bm{F}^{l}_1(p^l_i)^T\cdot\bm{F}^{l}_2(q^l_k))},
\end{equation}
where $\bm{F}^{l}_1$ and $\bm{F}^{l}_2$ are different fully connected layers T denotes the transposition of a matrix. Considering the calculation process of the middle layer, the dimensions of $p^l$ and $q^l$ here are not necessarily three. Then we update $p^l_i$ by using another two fully connected layer networks $\bm{F}^{l}_3$ and $\bm{F}^{l}_4$ based on the attention score and the point set $Q^l$ to make the value of $p^l_i$ geometrically more reliable. The update function is defined as
\begin{equation}
p^l_i \gets \bm{F}^{l}_4(p^l_i + \sum\limits _{q^l_j\in Q^l} a^l_{ij}\cdot\bm{F}^{l}_3(q^l_j)).
\end{equation}
When $P^l$ equals $Q^l$, AGB becomes a self-attention network, integrating similar internal features. The purpose of using multiple different fully connected layer networks in AGB is to establish connections between features in different domains (or the same). Compared with the generative model without AGB, the addition of AGB can significantly increase the expression ability in details, while also reducing the generation error and enhancing the stability of the model. We will prove this argument next.

\begin{figure*}[ht]
\centering
\includegraphics[width=18cm]{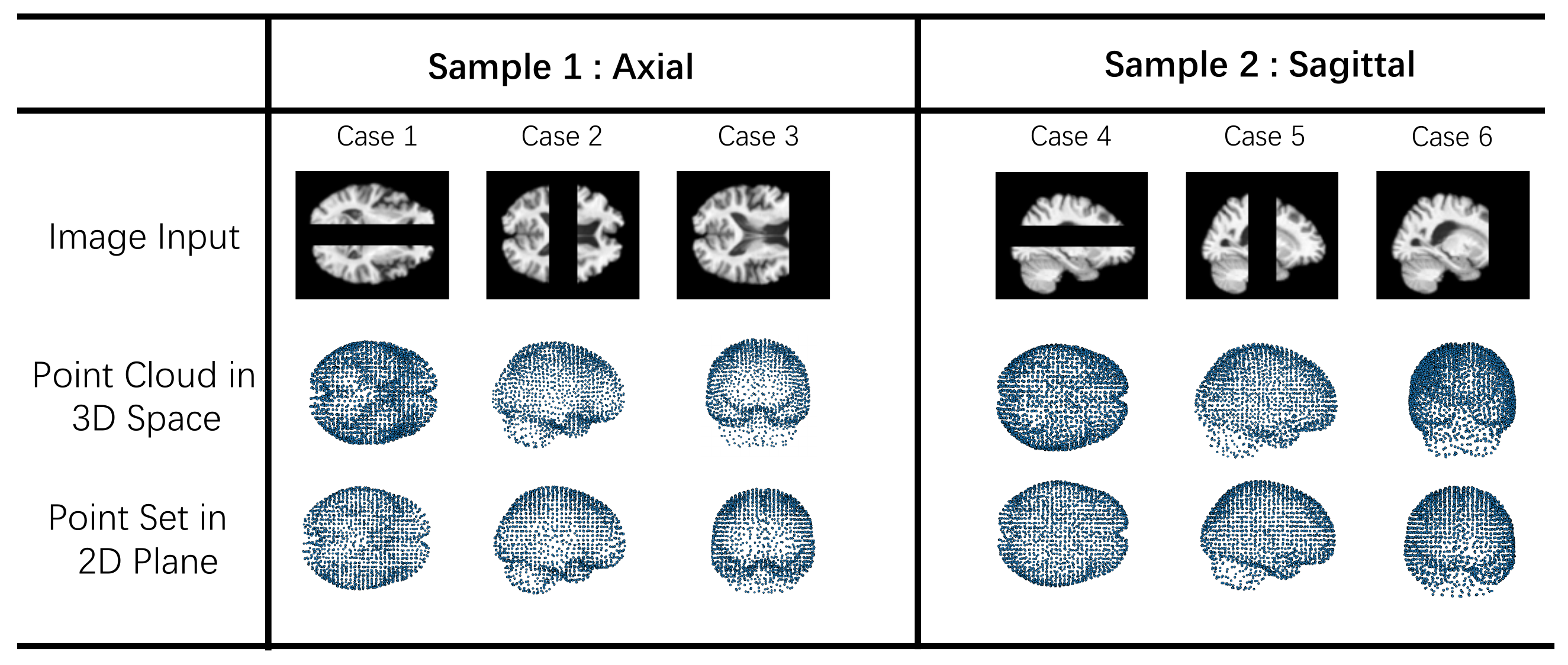}
\caption{Qualitative evaluations of the proposed model. We choose different angles of MRIs as inputs for the HSPN. We report the reconstructed point clouds by showing them in the 3D space and projecting them onto the 2D plane}
\end{figure*}

\subsection{Loss Functions in Training}
The proposed model consists of several different modules, working for different purposes. We have designed multiple loss functions to meet complex requirements. Because of the particular data format, conventional loss cannot train this generative network well. Chamfer distance is often used as loss function in point cloud generation and analysis. Given two point clouds $Y$ and $Y'$, Chamfer distance (CD) is defined as
\begin{equation}
\begin{aligned}
\mathcal{L}_{CD} = \sum\limits_{y' \in Y'} min_{y \in Y}& ||y' - y||^2_2 \\
& +  \sum\limits_{y \in Y} min_{y' \in Y'} ||y - y'||^2_2,
\end{aligned}
\end{equation}
where $y$ and $y'$ are points in $Y$ and $Y'$, respectively. We define the loss function of the ResNet and generator of predictor as
\begin{equation}
\mathcal{L}_{1} = \lambda_1 \mathcal{L}_{KL} + \lambda_2 \mathcal{L}_{CD} - \mathbb{E}_{z\sim \mathcal{Z}}[D(G(z))],
\end{equation}
where $ \mathcal{L}_{KL}$ is the Kullback-Leibler divergence, $\lambda_1$ and  $\lambda_2$ are variable parameters and $\mathcal{Z}$ is a Gaussian distribution calculated by encoder.

Meanwhile, the loss proposed by [12] is used on the discriminator. $\mathcal{L}_{2}$ is defined as
\begin{equation}
\begin{aligned}
\mathcal{L}_{2} =  \mathbb{E}_{z\sim \mathcal{Z}}[D(G(z))]& - \mathbb{E}_{Y\sim \mathcal{R}}[D(Y))]\\
& + \lambda_{gp}\mathbb{E}_{\hat{x}}[(||\nabla_{\hat{x}}D(\hat{x})||_2 - 1)^2],
\end{aligned}
\end{equation}
where $\hat{x}$ are sampled from line segments between real and fake point clouds, $\mathcal{R}$ represents the real point cloud distribution and $\lambda_{gp}$ is a weighting parameter.

For encoding blocks and the corresponding decoding blocks, Chamfer distance and another permutation-invariant metric for comparing unordered point clouds, Earth Mover's distance (EMD), are both used as the loss function. Given two point clouds $Y$ and $Y'$, Earth Mover's distance is defined as
\begin{equation}
\mathcal{L}_{EMD} = min_{\phi: Y \rightarrow Y'} \sum\limits_{x \in Y}||x - \phi(x)||_2
\end{equation}
where $\phi$ is a bijection. CD measures the distance between each point in one point cloud to its nearest neighbor in the other point cloud and helps points in the reconstruction point cloud closer to the corresponding position in the ground truth, while EMD is a measure of the transportation problem of trying to convert one point cloud to another point cloud and helps two point clouds to approach in the overall probability space. We build 3D joint perception loss to ensure the fitting ability of the complement network. The loss can be written as
\begin{equation}
\mathcal{L}_{3} = \lambda_3\mathcal{L}_{CD} + \lambda_4\mathcal{L}_{EMD}.
\end{equation}

\section{Experiment}
\begin{figure}[ht]
\includegraphics[width=8.8cm]{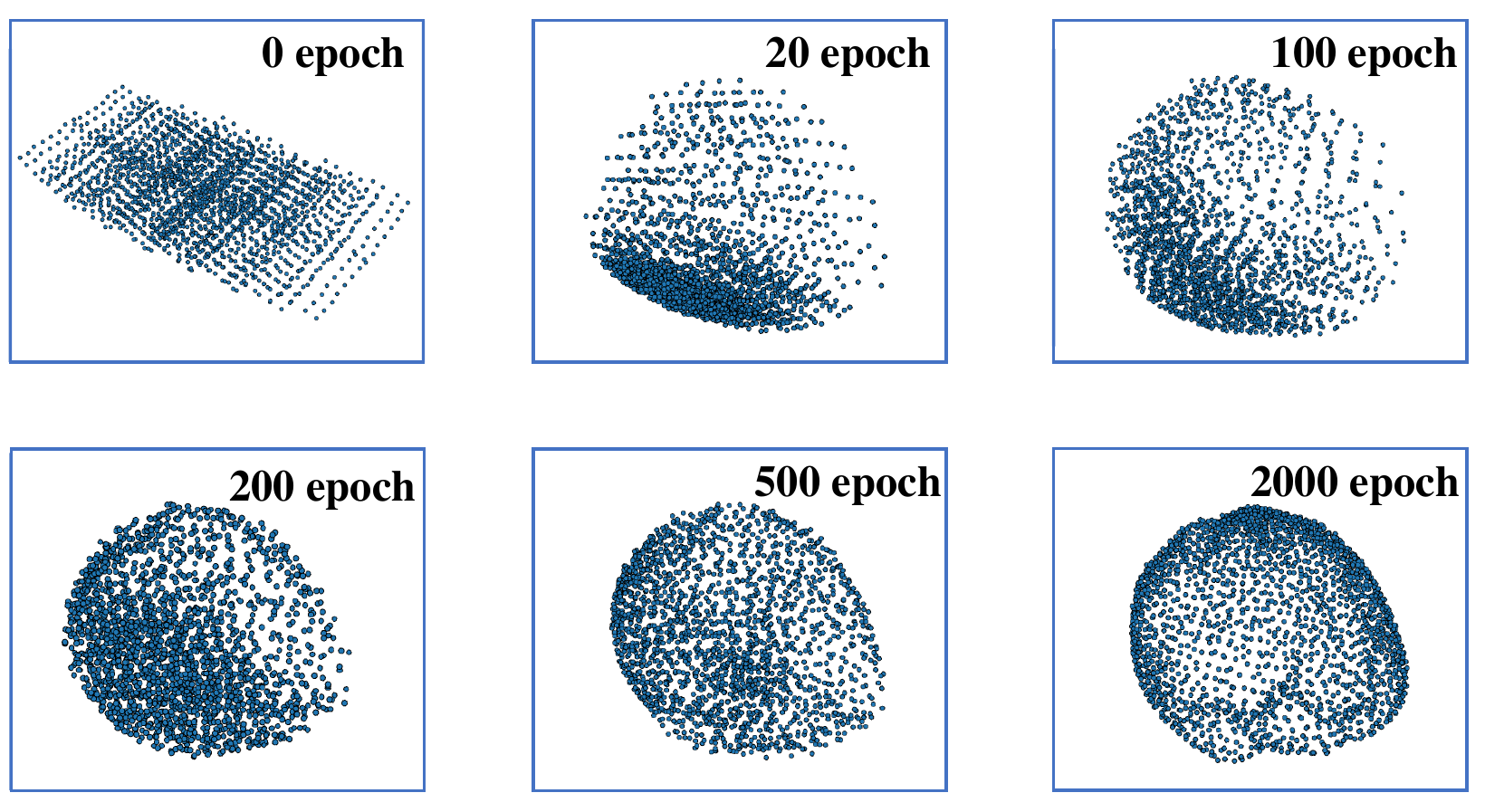}
\caption{The 3D reconstruction of brains during the training process.}
\end{figure}

\begin{figure*}[ht]
\centering
\includegraphics[width=18cm]{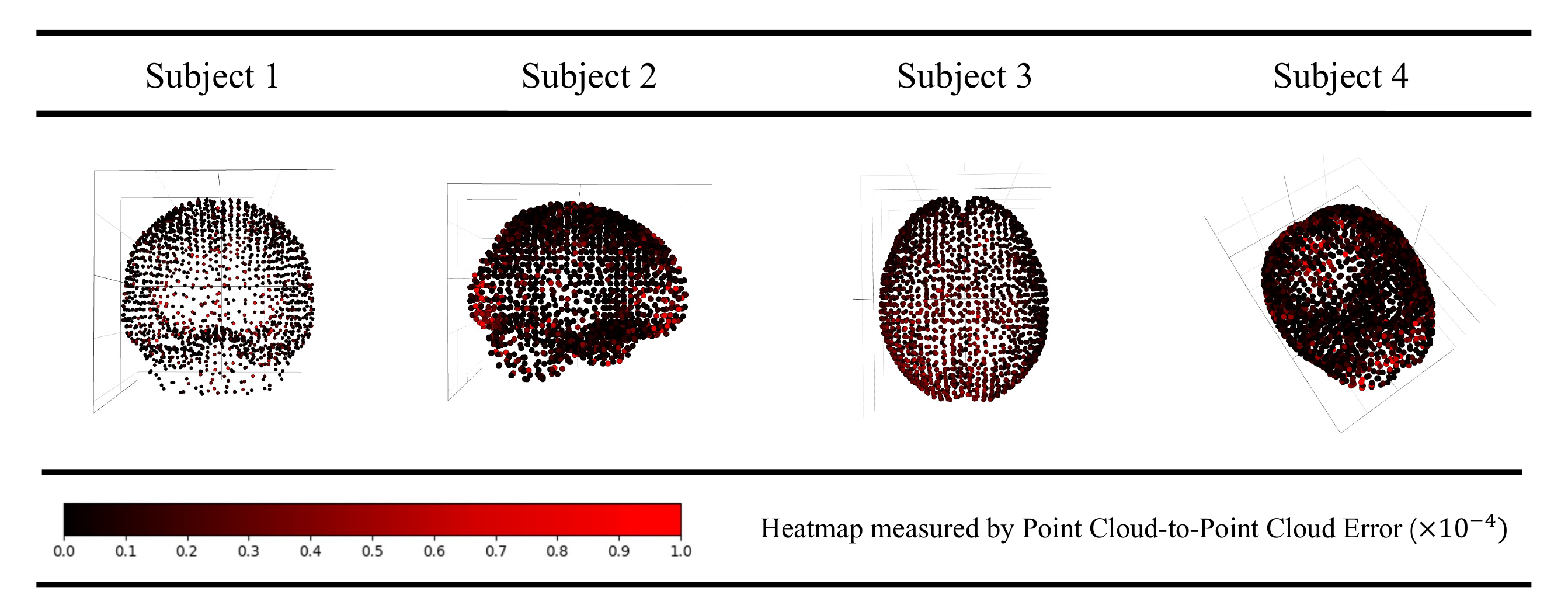}
\caption{The point-to-point reconstruction accuracy results in multiple of angles which is represented by the heat map colored by PC-to-PC error.}
\end{figure*}

\subsection{Data Preparation and Implementation Details}

To evaluate the performance of the proposed model, we conduct experiments on an in-house dataset. The preprocessing was prepared under the professional guidance of doctors. The dataset is composed of 317 brain MRIs with Alzheimer disease (AD) and 723 healthy brain MRIs. All bone structures of MRIs are removed, and the remaining images are registered into format 91 × 109 × 91 by the FSL software. 900 MRIs are randomly selected to construct the training set, and the others are used for the test set. We choose some 2D slices of MRIs which were normalized to [0, 1] as the input of the model. We build the origin dataset by performing accurate voxel-level segmentation and then converting it into point clouds.

2000 epochs were trained for our generative model and all the comparative models. HSPN is implemented with Pytorch, and the experiments are conducted on a CPU of Intel Core i9-7960X CPU @ 2.80GHz$\times$32 and GPUs of Nvidia GeForce RTX 2080 Ti. We set $\lambda_1$, $\lambda_3$, $\lambda_4$ and $\lambda_{gp}$ to 0.1, 1, 0.05 and 10, respectively, and use Adam optimisers with an initial learning rate of $1 \times 10^{-4}$. Specifically, we adjust training parameters dynamically by increasing $\lambda_2$ from 0.1 to 1.

In order to assess the effectiveness of the proposed model, Chamfer distance is utilized to assess the objective quality. Point Cloud-to-Point Cloud (PC-to-PC) error \cite{35} is utilized to measure the error value of each point in outputs. Besides, the robustness of the model is measured.

We conducted comprehensive analyses of HSPN on the built dataset. Considering the complexity of the task completed by the proposed model, we will prove the superiority of our model by replacing a part of the model with other existing high-performance network structures in ablation studies. In this section, we only show the reconstruction results of the proposed model.

\subsection{Evaluation of Completion Performance}

Fig. 4 reports some qualitative evaluations and their experimental results of HSPN on the test set. Fig. 5 shows the outputs of the proposed model in different training stages. Fig. 6 shows the point-by-point error of our generated point cloud. To further examine the detail capture ability, difference heatmap, in which each vertex is colored by PC-to-PC error, is used to reflects the absolute difference between the reconstructed point clouds and the corresponding ground truth. We select multiple angles the point clouds to present the effect of the heat map to fully illustrate the point-to-point reconstruction accuracy. The results show that most of the vertices in the reconstruction results have a very small error, and confirm the detailed sensitivity of the proposed structure. The value on the axis is multiplied by a factor $10^{-4}$.

\begin{figure*}[ht]
\centering
\includegraphics[width=15cm]{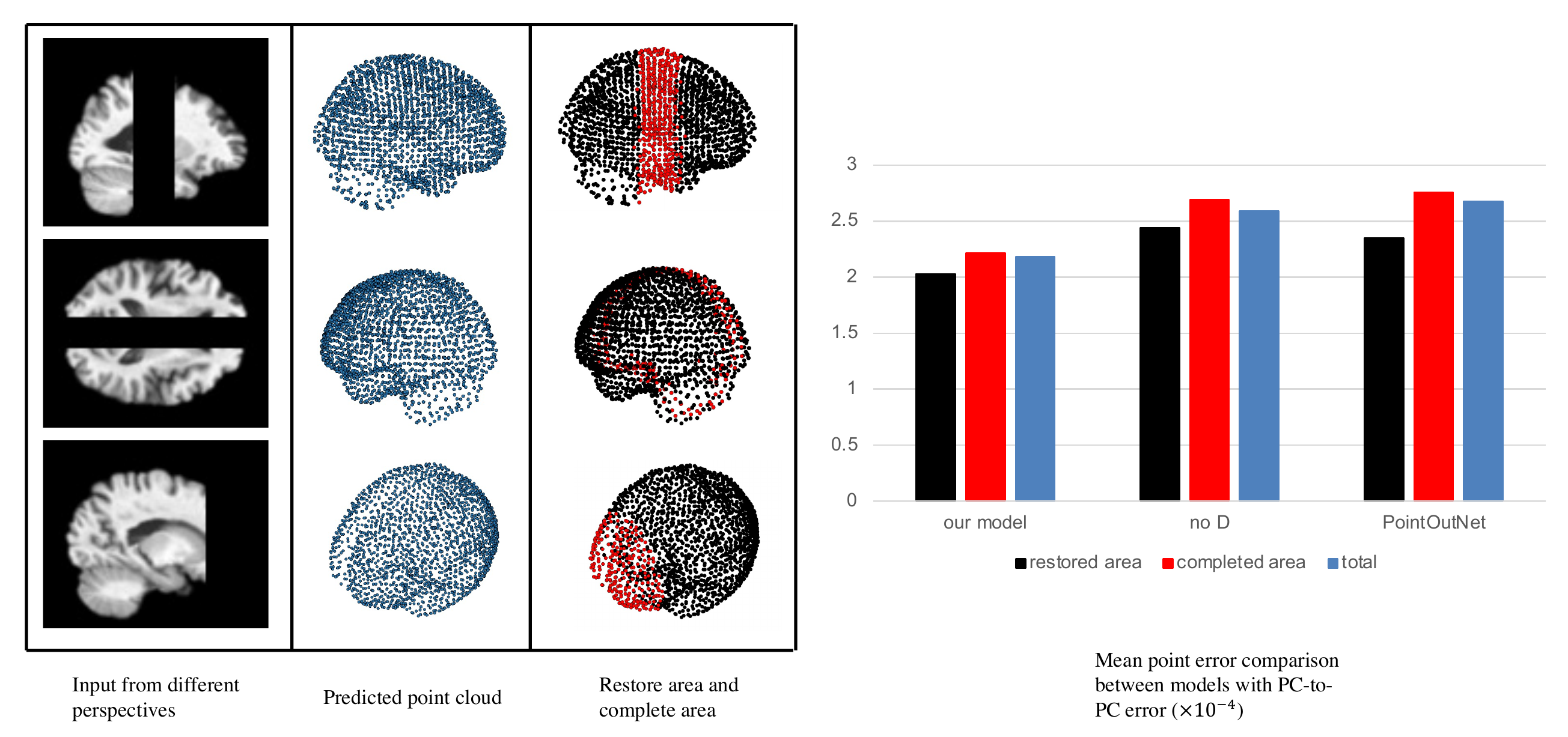}
\caption{Comparison results of different encoder structures. We report the reconstruction of complete area and incomplete area in the figure on the left and the average point cloud-to-point cloud error of the entire test in the figure on the right.}
\end{figure*}
\begin{figure*}[ht]
\centering
\includegraphics[width=15cm]{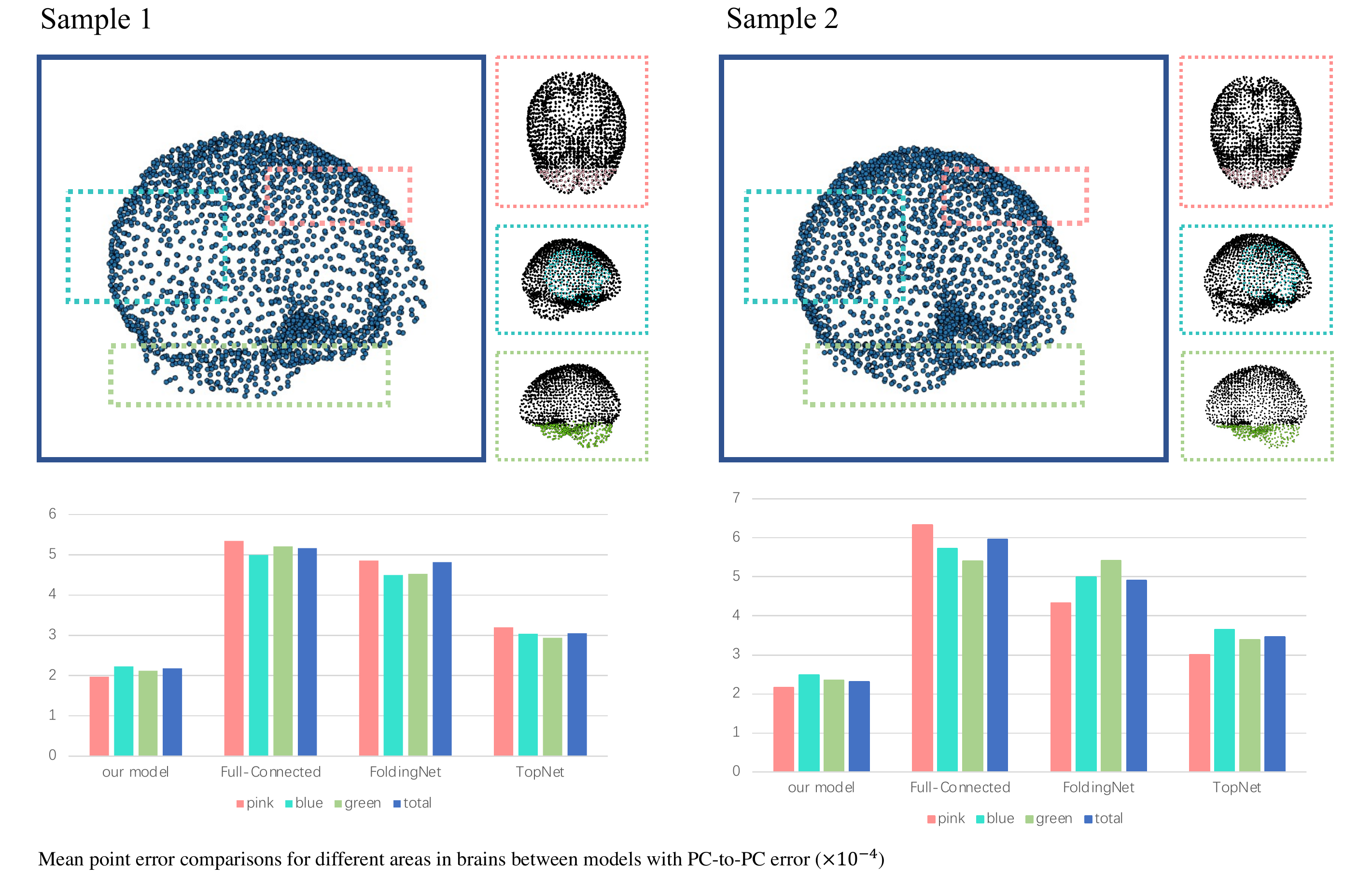}
\caption{Comparison results of different decoder structures. We report the sampling process in the figure on the left and the average point cloud-to-point cloud error of the entire test in the figure on the right.}
\end{figure*}

\subsection{Ablation Study}
In this subsection, we analyzed the effect of important modules and hyper-parameters to HSPN. All studies are typically conducted on the brain category for convenience. Considering that the number of generated points in each area may be different, we use the PC-to-PC error which is calculated for each point in point-error based evaluations.
2
\subsubsection{Effect of Predictor}
In order to prove the effectiveness and accuracy of our adversarial GCN predictor, we made the following two modifications to the predictor respectively and observed the error: (1) ``No-D" is the variation that removes the discriminator of the predictor, therefore canceling the adversarial generating principle of the predictor. In this case, we only use Chamfer distance as the loss function to train the ResNet and the generator. (2) ``PointOutNet" is the variation that uses another point cloud generation architecture PointOutNet in \cite{34} to replace our predictor structure. The comparison is shown in the left part of Fig. 7. The mean and overall point error comparison results are shown in the right part of Fig. 7 and Table I, respectively, which proves that the proposed structure can best help the model achieve accurate reconstruction targets.
\begin{table}[htbp]
	\centering
	\caption{The effect of different predictor structure measured by CD$(\times10^{-1})$}
	\begin{tabular}{p{2cm}|ccc}
		\toprule
		Method&Our Model&No-D&PointOutNet \\
		\midrule
		CD$(\times10^{-1})$&\color{red}{4.461}&5.309&\color{blue}{5.492} \\
		\bottomrule
	\end{tabular}
\end{table}

\subsubsection{Effect of Completion System}
The generating ability of the completion system directly affects the fitting effect of the model and determines the final quality of completed point clouds. We used several different models to replace the encoder (if need) and decoder we built and tested their performance: (1) ``FC" is the variation that conducts a fully-connected decoder to replace the proposed decoder framework. the size of inputs and outputs in the three-layer network is respectively $[96,1024], [1024,2048], [2048,2048\times3]$. (2) ``FN" is the variation that conducts a FoldingNet based encoder and decoder proposed by \cite{28}. FoldingNet is usually a generative model, but it can also be generalized to point cloud completion tasks by changing the inputs appropriately. (3) ``TN" is to replace encoding blocks and decoding blocks with TopNet in \cite{39}. We show the process of two samples which are selected randomly of the test in the top part of Fig. 8 and report results in the bottom part of Fig. 8 and Table II. The mean and total point error comparisons demonstrate the strongest reconstruction ability of our model.

\begin{table}[htbp]
	\centering
	\caption{The effect of completion system to the model measured by CD$(\times10^{-1})$}
	\begin{tabular}{p{2cm}|cccc}
		\toprule
		Method&Our Model&FC&FN&TN \\
		\midrule
		CD$(\times10^{-1})$&\color{red}{4.461}&\color{blue}{10.572}&9.863&6.255 \\
		\bottomrule
	\end{tabular}
\end{table}

\begin{figure*}[ht]
\centering
\includegraphics[width=15cm]{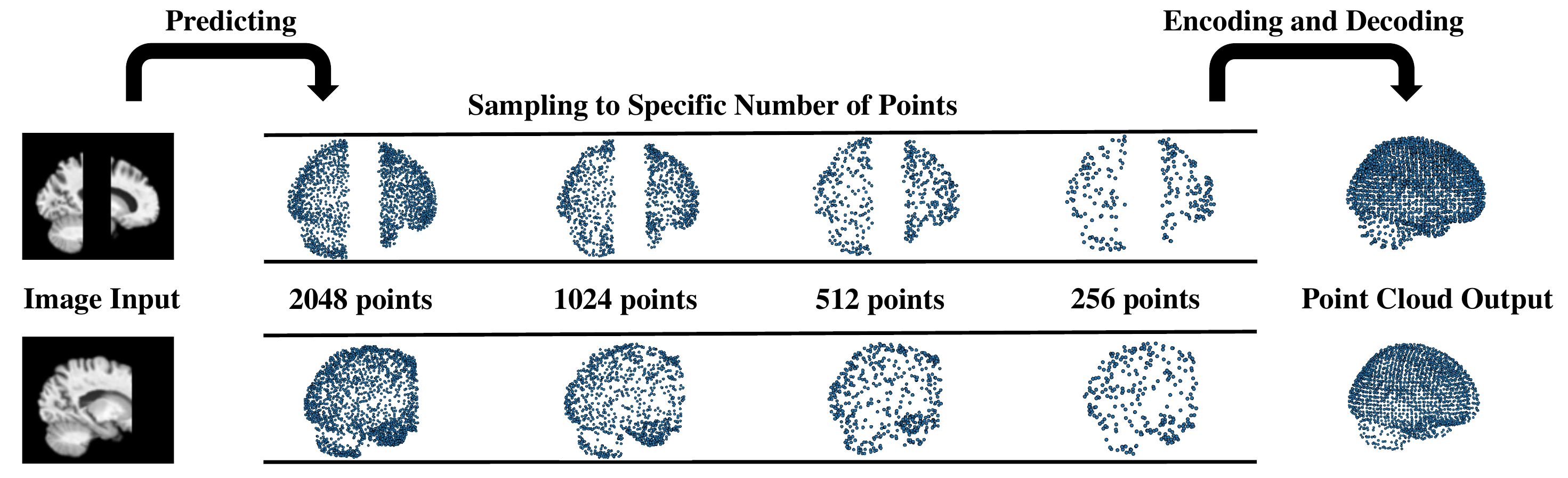}
\caption{Comparison process of different point nums of the predictor. We randomly sample points according to a predetermined number from the results of 2048 points and input them into the subsequent module.}
\end{figure*}

\subsubsection{Effect of AGB}
We developed four variations of the proposed model to verify the effectiveness of the attention gate blocks: (1) The first line of Table III: the variation that removes all the AGBs from the model. (2) The second line of Table III: the variation that removes hierarchical attention pipelines and the corresponding AGBs $(AGB^1)$. (3) The third line of Table III: the variation that removes the AGBs which work as the self-attention blocks $(AGB^2)$. (3) The last line of Table III: the variation that remains all the AGBs in the model. All four variations have the same structure except for the removed/replaced module. The results are shown in Table III, in which the model that remains all AGBs achieves the best performance. The experimental results prove the effectiveness of the attention gate block used in our model.

\begin{table}[h]
	\centering
	\caption{The effect of AGB to our model, all results are measured by CD$(\times10^{-1})$}
	\begin{tabular}{p{0.6cm}p{0.8cm}|ccc}
		\toprule
		$AGB^1$&$AGB^2$&1000th epoch&1500th epoch&2000 epoch\\
		\midrule
		\XSolidBrush&\XSolidBrush&5.258&5.071&4.958 \\
		\XSolidBrush&\Checkmark&5.160&4.952&4.831\\
		\Checkmark&\XSolidBrush&\color{blue}{5.314}&\color{blue}{5.186}&\color{blue}{5.178}\\
		\Checkmark&\Checkmark&\color{red}{4.741}&\color{red}{4.406}&\color{red}{4.461}\\
		\bottomrule
	\end{tabular}
\end{table}

\subsubsection{Effect of Point Number measured by the total point error}
In this work, the size of the point cloud output of our predictor and decoder is 2048, and the model performs well on this scale. And we also analyze whether the model still performs similarly when the input conditions deteriorate. We test the robustness of decoder by randomly select n points of outputs of the predictor and transfer them to the first encoding block, so as to ensure that the generation efficiency will not be affected by the visual and sensing capabilities during the surgery. We set mutiple of n in the test which is explained by Fig. 9 and the results are shown in Table IV.

\begin{table}[htbp]
	\centering
	\caption{The effect of point number measured by CD$(\times10^{-1})$}
	\begin{tabular}{p{1.5cm}|cccc}
		\toprule
		Method&2048 Points&1024 Points&512 Points&256 Points \\
		\midrule
		CD$(\times10^{-1})$&\color{red}{4.461}&4.655&4.836&\color{blue}{5.178} \\
		\bottomrule
	\end{tabular}
\end{table}

\subsubsection{Effect of Image Input Number}
In this work, we use a single image to reconstruct the 3D point cloud instance of the target brain and proved the high accuracy of the reconstructed point cloud. Meanwhile, we conjecture that since the MRI image contains structural information corresponding to the human brain, using multiple MRI slices instead of a single image as the input of the model may improve the quality of the features aggregated by the encoder, thereby improving the reconstruction quality of the decoder. Therefore, in this section, we design to use different numbers of 2D MRIs instead of single image input and calculate the output errors respectively. The results are reported in Table V. It can be seen that although the input of multiple slices does improve the accuracy of the reconstruction, the degree of improvement is not obvious, and processing multiple images will significantly increase the computational complexity of the model. Therefore, we believe that a single image input is appropriate for this work.
\begin{table}[htbp]
	\centering
	\caption{The effect of image input number measured by CD$(\times10^{-1})$}
	\begin{tabular}{p{1.5cm}|cccc}
		\toprule
		Method&single slice&3 slices&5 slices&7 slices \\
		\midrule
		CD$(\times10^{-1})$&4.461&4.327&\textbf{4.285}&4.369 \\
		\bottomrule
	\end{tabular}
\end{table}

\subsection{Classification Experiment on the Reconstruction Point Clouds}
\begin{figure}[h]
\centering
\includegraphics[width=8.8cm]{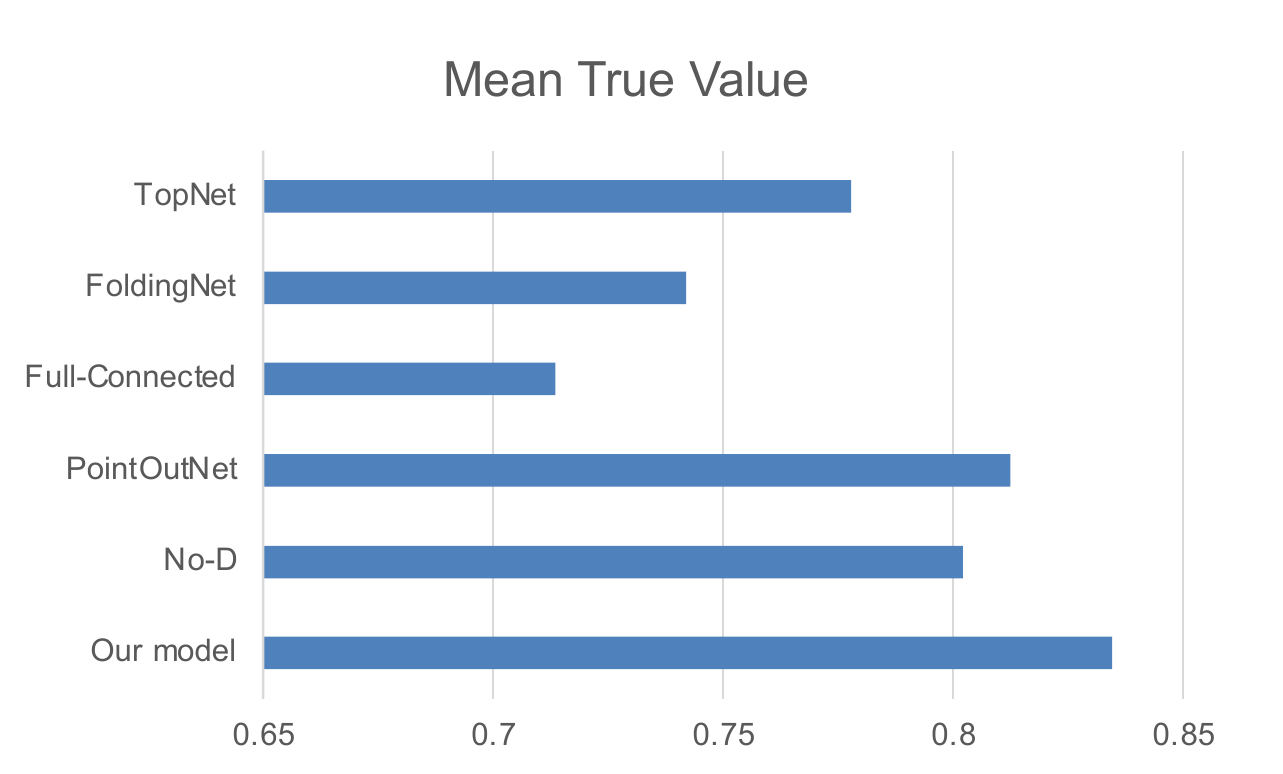}
\caption{The result of classification experiment on the reconstruction point clouds}
\end{figure}
Nevertheless, Neither CD nor EMD can show the visual reconstruction quality of outputs directly. To study the generative effect of point clouds in different models, we further observed the classification accuracy of the reconstruction point clouds which are generated by the proposed method and all the methods in the ablation study on a binary classification task with pointNet++ classifier network. For a fair comparison, we use the idea of training the discriminator in GANs to train our classifier. The pointNet++ is pre-trained first by using the point cloud ground truth (labeled as true) and point clouds which are generated by each method trained to 1000 epochs (labeled as false). Then, the generated point clouds from different methods in the test were input to the classifier, respectively. We compare the ability of methods to fool this classifier through such a classification experiment. The results of the experiment are shown in Fig. 10. A higher average true value means that the classifier is more confident that point clouds generated by a method are real point clouds. It can be observed that the ability for fooling the classifier of the proposed model outperformed the other compared methods. As is shown in these results, we can make a conclusion that the point clouds generated by the proposed architecture have the best quality and the most realistic details, and restores the most microstructure information of the actual brain structure.

\section{Discussion and Conclusion}
In this paper, we propose a novel image-to-PC reconstruction model named hierarchical shape-perception network to tackle the problem of visual occlusion and incompleteness in minimally-invasive and robot-guided surgeries, so as to enhance the versatility and robustness of shape reconstruction technology in surgical scenarios. Considering the real-time feedback requirement during surgery and the computational complexity, the point cloud is used as a representation of the reconstruction network, and a single incomplete image is used as the input.

A naive alternative to our method would be to train a simple generative network and predict the whole point cloud structure from the broken image in one step. However, a direct generative model is infeasible because a) the known information in the image and prior knowledge are not fully utilized and b) the complete area to be restored and the damaged area to be completed in the image are not reconstructed prudently, both of which will cause a great increase in reconstruction error.

Our multi-structure model integrates multiple modules to jointly complete scheduled complex prediction and completion tasks. An adversarial branching predictor is designed to extract the rich structural information in 2D MRI and predict the initial incomplete point clouds. Hierarchical encoding and decoding blocks are used to fulfill the completion task while ensuring the perception of microstructure. General logic AGBs are constructed, can output trainable results of multiple attentions based on the same mechanism.

At present, intraoperative MRI technology has been developed and has made great progress in brain surgery. In this paper, we use pre-processed 2D MRIs for experiments and have achieved good results. We hope that this work can immediately help doctors obtain the 3D shape visual and positional information of brains during surgery. In the long-term work plan, we aim to build an intraoperative real-time internal organs image dataset.  We hope to reduce the input constraints of HSPN as much as possible to eliminate the unavailability of the model in more scenarios and time.

Considering that almost no work of the current work is completely consistent with the purpose and operating mode of HSPN, we prove the advantages and effectiveness of our model by replacing different parts of the model composition with existing advanced work and then comparing. In qualitative and quantitative experiments, our model behaves better than the other state-of-the-art combination. Our method has a significantly short inference time, which enables effective real-time feedback of local image properties. This feedback can guide doctors to find diagnostically valuable surgical locations.


%



\section*{Acknowledgment}
This work was supported by the National Natural Science Foundations of China under Grant 61872351, the International Science and Technology Cooperation Projects of Guangdong under Grant 2019A050510030, the Distinguished Young Scholars Fund of Guangdong under Grant 2021B1515020019, the Excellent Young Scholars of Shenzhen under Grant RCYX20200714114641211 and Shenzhen Key Basic Research Project under Grant JCYJ20200109115641762.




\bibliographystyle{IEEEtran}
\bibliography{bare_jrnl}

\end{document}